\documentclass[showpacs,fleqn,nobibnotes]{revtex4}

\usepackage{amsmath}
\usepackage{graphicx}
\usepackage{float}

\def\lsim{\raise0.3ex\hbox{$<$\kern-0.75em\raise-1.1ex\hbox{$\sim$}}}
\def\gsim{\raise0.3ex\hbox{$>$\kern-0.75em\raise-1.1ex\hbox{$\sim$}}}

\newcommand{\rr}{\mbox{\boldmath $r$}}

\newcommand{\rb}{\mbox{\boldmath $b$}}

\newcommand{\be}{\begin{equation}}
\newcommand{\ee}{\end{equation}}

\begin{document}

\title{Heavy quark production at LHC in the color dipole formalism}
\author{E.R.  Cazaroto$^1$ , V.P. Gon\c{c}alves$^2$ and F.S. Navarra$^1$}

\affiliation{$^1$ Instituto de F\'{\i}sica, Universidade de S\~{a}o Paulo,
C.P. 66318,  05315-970 S\~{a}o Paulo, SP, Brazil\\
 $^2$ Instituto de F\'{\i}sica e Matem\'atica, Universidade Federal de
Pelotas\\
Caixa Postal 354, CEP 96010-900, Pelotas, RS, Brazil.}

\begin{abstract}

In this work we estimate the heavy quark production in proton-proton and 
proton-nucleus collisions at LHC energies 
using the color dipole formalism and the solution of the running coupling 
Balitsky-Kovchegov equation. Nuclear effects 
are considered in the computation of the total cross sections and rapidity 
distributions for scattering on protons and  
nuclei.
 
\end{abstract}

\maketitle

\section{Introduction}

Heavy quark production in high energy collisions is important for a number of reasons. 
One of them is that it offers a 
good testing ground for perturbative QCD (pQCD) calculations, which in spite of the 
continuous progress over the last 
thirty years, still contain  some ambiguities, mostly related to scale dependence of
 the observable quantities.  In the 
context of pQCD there are different calculation schemes. The most well known is the   
collinear factorization approach, 
which has been developed up to next-to-leading order (NLO) in $\alpha_s$ 
\cite{nason0,alta,nason1,been}. More recently the  
Fixed Order plus Next-to-Leading-Log (FONLL) \cite{greco} was developed.
It  resumes large perturbative terms proportional to powers of    
$\alpha_s \,  log (p_T/m_Q)$  where $p_T$ and  $m_Q$ are the  
heavy quark transverse  momentum and  mass, respectively. 
In these calculations all particles involved are assumed to be on mass 
shell, carrying only longitudinal momenta, and the cross section is averaged over 
two transverse polarizations of the incident 
gluons. The virtualities $Q^2$ of the initial partons are taken into account only 
through their structure functions.
The formalism which incorporates the incident parton transverse momenta is referred
to as the $k_T$-factorization approach \cite{cata90,cata91,collins91,marchesini88,hagler,
gribov83,levin90,shab2004}.  
In this approach  the Feynman diagrams are calculated taking into account the 
virtualities and all possible polarizations of the incident partons.  In the $k_T$-factorization 
approach the unintegrated gluon distributions are 
used instead of the usual structure functions.

In  proton nucleus collisions the linear A-dependence of the heavy 
flavor production cross sections is usually assumed, 
so the obtained results are  discussed
in terms of the cross section scaled to pp collision.  
In perturbative QCD with the factorization approximation,   heavy ion 
collisions differ from p-p collisions only by the change of the usual nucleon structure functions 
or unintegrated gluon distributions by the same functions for bound nucleons. 
At RHIC energies the difference between bound and unbound  parton structure functions is known from EMC and 
NMC experimental data  and their DGLAP evolution  analysis.  From these studies we have learned that 
nuclear effects in  charm production at RHIC 
cause a deviation  of $5 - 10$  \%  from the linear (A-dependent) predictions.
The experimental data of the PHENIX \cite{adcox02,adare06}  collaboration obtained at RHIC 
both from pp and nuclear collisions are in reasonable agreement with the NLO QCD and with the
$k_T$  factorization approach predictions. The existing  STAR  \cite{biel06} 
collaboration data are in contradiction with the PHENIX data and with standard QCD calculations. However 
a reanalysis of these data is in progress.

Calculations performed with other approaches, also based on pQCD,  are welcome and 
may be very useful to cross-check and complement 
previous analyses. In this paper we calculate  heavy quark production cross sections using a formalism 
derived from the high energy regime of QCD (For recent reviews see Ref. \cite{hdqcd}). In
this regime, perturbative QCD  predicts that
the small-$x$ gluons in a hadron wavefunction should form a Color
Glass Condensate (CGC), which  is characterized by the limitation on the maximum
phase-space parton density that can be reached in the hadron
wavefunction (parton saturation), with the transition being
specified  by a typical scale, which is energy dependent and is
called saturation scale $Q_{\mathrm{sat}}$. In order to estimate the
saturation effects we study  heavy quark production using the
color dipole approach, which gives a simple and unified picture of  this
process in photon-hadron and hadron-hadron interactions. It is
important to emphasize that in contrast to the heavy quark
production in the previous accelerators (SPS, Tevatron and RHIC), where the saturation scale
$Q_{\mathrm{sat}}$ is smaller than the typical hard scale, $\mu =
 m_Q$, at the LHC energies,  we  probe
 for the first time the kinematical regime where $Q_{\mathrm{sat}} \approx
 \mu$. Therefore, at these energies  one may expect a sizeable 
 modification of the heavy quark total cross sections and rapidity distributions.

This paper is organized as follows. In next section (Section \ref{cross}) we present a 
brief review of the heavy quark production 
in the color dipole formalism, introducing the main formulae. In Section \ref{dinamica} 
we discuss the QCD dynamics and the models 
used in the calculations. In Section \ref{resultados} we  present our predictions for 
the total cross sections and rapidity 
distributions for heavy quark production in $pp$ and $pA$ collisions. Finally, in 
Section \ref{conc} we summarize our main results 
and conclusions.

\section{Heavy quark production in the color dipole formalism}
\label{cross}

Heavy quark hadroproduction  at high energies has been usually described
considering the collinear factorization, where  all partons involved are assumed to 
be on mass shell, carrying 
only longitudinal momenta, and their transverse momenta are neglected in the QCD 
matrix elements. The cross sections  
of the QCD subprocess are usually calculated in leading order (LO), as well as in  
next-to-leading order (NLO). 
In particular, the cross sections involving incoming hadrons are given, at all orders, 
by the convolution of intrinsically 
non-perturbative (but universal) quantities - the parton densities - with perturbatively calculable hard matrix elements, 
which are process dependent. The conventional gluon distribution $G(x,\mu^2)$, which drives the behavior of the observables 
at high energies, corresponds to the density of gluons in the proton having a longitudinal momentum fraction $x$ at the 
factorization scale $\mu$. This distribution satisfies the DGLAP evolution in $\mu^2$ and does not contain information about 
the transverse momentuum $k_T$ of the gluon. On the other hand, in the large energy (small-$x$) limit, the characteristic scale 
$\mu$ of the hard subprocess of parton scattering is much less than $\sqrt{s}$, but greater than the $\Lambda_{QCD}$ parameter. 
In this limit, the effects of the finite transverse momenta of the incoming partons become important, and the factorization must 
be generalized, implying that the cross sections are now $k_T$-factorized into an off-shell partonic cross section and a  
$k_T$-unintegrated parton density function ${\cal{F}}(x,k_T)$, characterizing the $k_T$-factorization  approach.  
The function $\cal{F}$ is obtained as a solution  of the   evolution equation associated to the dynamics that governs  QCD at high 
energies. Several authors have considered the $k_T$-factorization approach in order to  analyse  some  observables and 
they have obtained a better description of these quantities than the one obtained with  the collinear approach.    
However, the current situation is still not satisfactory, due to the large uncertainty   associated to the lack of a complete knowledge 
of the unintegrated gluon distribution. In particular, at high energies the non-linear QCD effects associated to the gluon saturation are 
expected to contribute significantly and this leads to the breakdown of the twist expansion and of the factorization schemes.

Gluon saturation effects can be naturally described in the color dipole formalism. At high energies color dipoles with a defined 
transverse 
separation are  eigenstates of the interaction. The main quantity in this formalism is the dipole-target cross section, which is 
universal and  
determined by  QCD dynamics at high energies. In particular, it provides an unified description of inclusive and diffractive observables 
in $ep$ processes as well as for in  Drell-Yan, prompt photon and heavy quark production in hadron-hadron collisions. Furthermore, an important 
advantage of this formalism is that it is very simple  to include nuclear effects (See below).

In the color dipole formalism the  total cross section  for the process $h_1h_2 \rightarrow Q\bar{Q} X$, where $h_i = p$ or $A$, 
is given by \cite{rauf}:
\begin{equation}
\sigma _{tot} (h_1 h_2 \rightarrow \{ Q\bar{Q} \}X) = 
2 \int _0 ^{-ln(2m_Q/ \sqrt{s})} dy  \, x_1 \, G_{h_1}(x_1,\mu _F) \,
\sigma (Gh_2 \rightarrow \{Q\bar{Q}\} X)
\label{sigtot}
\end{equation}
where $x_1G_{h_1}(x_1,\mu _F)$ is the projectile gluon distribution, the cross section  $\sigma (Gh_2 \rightarrow \{Q\bar{Q}\} X)$ describes 
the heavy quark production in the gluon - target interaction, $y$  is the rapidity of the pair and $\mu_F$ is the 
factorization  scale, which we assume to be given by $\mu_F=2 m_Q$. The proton gluon distribution  will be taken from 
the set of parton densities GRV98 \cite{grv}, but similar predictions are obtained using e.g. the CTEQ6L parameterization.  Eq. (\ref{sigtot}) can be easily interpreted in the  target rest frame, where heavy quark
production looks like pair creation in the target color field.  For a short time, a gluon
G from the projectile hadron can develop a fluctuation which contains a heavy quark pair
($Q \bar{Q}$). The interaction with the color field of the target  may then release these heavy quarks.
This  mechanism  corresponds to the gluon-gluon fusion mechanism of heavy quark production in
the leading order (LO) parton model. The dipole formulation is therefore applicable only at 
low $x$, where the gluon density of the target is much larger than all quark densities.
This condition is fulfilled for the charm and bottom production at the CERN-LHC.

The cross section for the process $G + h_2 \rightarrow Q \bar{Q} X$ is given by:
\begin{equation}
\sigma(G h_2 \rightarrow\{Q\bar{Q}\}X) = \int _0^1 d \alpha \int d^2\rho \,\, 
\vert \Psi _{G\rightarrow Q\bar{Q}} (\alpha,\rho)\vert ^2 
\,\, \sigma^{h_2} _{q\bar{q}G}(\alpha , \rho)
\label{sec1}
\end{equation}
where $  \sigma^{h_2}_{q\bar{q}G}$  is the scattering cross section of a color neutral quark-antiquark-gluon system on the 
hadron target $h_2$ \cite{rauf}: 
\begin{equation}
\sigma^{h_2}_{q\bar{q}G}(\alpha , \rho) = \frac{9}{8}[\sigma _{q\bar{q}}(\alpha \rho) + \sigma _{q\bar{q}}(\bar{\alpha} \rho)]
- \frac{1}{8}\sigma _{q\bar{q}}(\rho)\,\,. 
\label{sec2}
\end{equation}
The quantity $\sigma _{q\bar{q}}$ is the scattering cross section of a  color neutral quark-antiquark
pair with separation radius $\rho$ on the target  and $\alpha$ ($\bar{\alpha} = 1 - \alpha$) is the 
fractional momentum of quark (antiquark). The light-cone (LC) wavefunction of the 
transition $G \rightarrow  Q \bar{Q} $ can be calculated perturbatively, with the squared wavefunction given by:
\begin{equation}
\vert \Psi _{G\rightarrow Q\bar{Q}} (\alpha,\rho)\vert ^2   
= \frac{\alpha _s (\mu _R)}{(2\pi)^2} \lbrace m^2_Q K_0^2(m_Q\rho)
+ [\alpha ^2 + \bar{\alpha ^2}] m^2_Q K_1^2(m_Q\rho)\rbrace
\label{psi}
\end{equation}
where $\alpha_s(\mu_R)$ is the strong coupling constant, which is computed at a 
renormalization scale $\mu_R$, which we assume to be equal to quark mass,  and it is given by:
\begin{equation}
\alpha _s (\mu _R) = \frac{4\pi}{ (11- \frac{2}{3}N_f) 
\, ln( \frac{\mu \,  _R ^2}{(200 \, MeV)^2} )}\,\,.
\label{alfa}
\end{equation}
Another observable of interest is the rapidity distribution, which in the dipole formalism is expressed as:
\begin{equation}
\frac{d\sigma(h_1 h_2 \rightarrow \{ Q\bar{Q} \}X)}{dy} = x_1G^{h_1}(x_1,\mu_F^2) \, 
\sigma (Gh_2 \rightarrow \{Q\bar{Q} \}X)\,\,.
\label{dsdy}
\end{equation}
Before discussing the QCD dynamics at high energies in the next section, some comments are in order. First, in the dipole formalism we 
work in a mixed representation, where the longitudinal direction is treated in momentum space, while the transverse directions are described 
in the coordinate space representation. Second, in Ref. \cite{rauf} the equivalence between this
approach and the gluon-gluon fusion mechanism of the conventional  collinear factorization  at leading order and twist has been demonstrated. 
In particular, the  dipole predictions are similar to those obtained using  the
 next-to-leading order parton model calculation. However,  it is important to emphasize that Eq. (\ref{sigtot}) resums higher-twist  corrections 
beyond the traditional factorization schemes.

\section{QCD dynamics}
\label{dinamica}

The main ingredient of the dipole formalism is the dipole-target cross section $\sigma _{q\bar{q}}$ which is determined by the 
QCD dynamics. At leading logarithmic approximation it is  directly related to the target gluon distribution $x \, G_{h_2}$ as 
follows  \cite{nik}: 
\begin{eqnarray}
\sigma_{q\bar{q}}(x,\rho^2) =  {\frac{\pi^2}{3}} \rr^2 \alpha_s \, x \, G_{h_2}(x, 10/\rho^2) \,\,,
\label{gluongrv}
\end{eqnarray}
which satisfies the property known as color transparency, i.e. $\sigma_{q\bar{q}}$ vanishes $\propto \rho^2$ at small separations.
If we assume that $x \, G_{h_2}$ is a solution of the DGLAP evolution equations, the use of this expression as input in our calculations implies 
that we are disregarding  non-linear QCD effects, associated to the high gluon density present at small-$x$ (large energies). In what follows 
we assume that the resulting predictions correspond to the linear QCD dynamics and denote them by CT in the plots.
When the target is a proton we assume that the   gluon distribution  is given by the GRV98 parametrization \cite{grv}. 
If the target is a nucleus we will assume that $x \, G_A (x,Q^2) = A . R_g(x,Q^2) . x \, G_N (x,Q^2)$ with  $x \, G_N$ being the  
gluon distribution in the proton  and $R_g$ given by the EPS09 \cite{eps09} or DS \cite{nds} parametrizations for the nuclear 
effects. In both cases  we are disregarding multiple 
scatterings of the dipole with the nuclei and are assuming that the dipole interacts incoherently with the target. Expression (\ref{gluongrv}) was  recently used in 
\cite{hqprc} to give the  linear physics predicitons for heavy quark production in 
$e \, A$ collisions.

The  results from HERA, RHIC and more recently from the LHC suggest that $ep$ and hadron - hadron interactions at high energies probe 
QCD in the non-linear regime of high parton densities,  where the growth of parton distributions should saturate, possibly forming a
Color Glass Condensate \cite{CGC}.  This formalism implies that the  dipole - target cross section  
$\sigma_{q \bar q}$ is given in terms of the dipole-target forward scattering amplitude ${\cal{N}}(x,\rho,\rb)$, which encodes all the
information about the hadronic scattering, and thus about the non-linear and quantum effects in the hadron wave function. It reads: 
\begin{eqnarray}
\sigma_{q \bar q} (x,\rho)=2 \int d^2 \rb \, {\cal{N}}(x,\rho,\rb)\,\,.
\end{eqnarray}
It is
useful to assume that the impact parameter dependence of $\cal{N}$ can be factorized as 
${\cal{N}}(x,\rho,\rb) = {\cal{N}}(x,\rho) S(\rb)$, so that 
$\sigma_{q \bar q}(x,\rho) = {\sigma_0} \,{\cal{N}}(x,\rho)$, 
with $\sigma_0$ being   a free parameter
related to the non-perturbative QCD physics. The Balitsky-JIMWLK hierarchy \cite{CGC}  describes the energy evolution of the dipole-target
scattering amplitude ${\cal{N}}(x,\rho)$. In the mean field approximation, the first equation of this  hierarchy decouples and boils down 
to the Balitsky-Kovchegov (BK) equation \cite{BAL,KOVCHEGOV}.

In the last years the next-to-leading order corrections to the  BK equation were
 calculated  
\cite{kovwei1,javier_kov,balnlo} through the ressumation of $\alpha_s N_f$ contributions to 
all orders, where $N_f$ is the number of flavors. Such calculation allows one to estimate 
the soft gluon emission and running coupling corrections to the evolution kernel.
The authors have found out  that  the dominant contributions come from the running 
coupling corrections, which allow us to  determine the scale of the running coupling in the 
kernel. The solution of the improved BK equation was studied in detail in Refs. 
\cite{javier_kov,javier_prl}. The running of the coupling reduces 
the speed of the evolution to values compatible with experimental data, with the geometric 
scaling regime being reached only at ultra-high energies. In \cite{bkrunning} a global 
analysis of the small $x$ data for the proton structure function using the improved BK 
equation was performed  (See also Ref. \cite{weigert}). In contrast to the  BK  equation 
at leading logarithmic $\alpha_s \ln (1/x)$ approximation, which  fails to describe the 
HERA data, the inclusion of running coupling effects in the evolution renders the BK equation 
compatible with them (See also \cite{vic_joao,alba_marquet,vicmagane}). In what follows we 
consider 
the BK predictions for ${\cal{N}}(x,\rho)$ obtained using the GBW \cite{dipolos} initial 
condition.

The dipole-target cross section  can also be calculated considering  phenomenological 
parametrizations 
for ${\cal{N}}(x,\rho)$ based on saturation physics, which provide 
an economical description of a wide range of data with a few parameters.
Several models for the forward dipole cross section have been used in the literature 
in order to fit the HERA and RHIC data. 
In general, the   dipole scattering amplitude is modelled in the coordinate space in terms 
of a simple Glauber-like formula as follows
\begin{eqnarray}
{\cal{N}}(x,\rho) = 1 - \exp\left[ -\frac{1}{4} 
(\rho^2 Q_s^2)^{\gamma (x,\rho^2)} \right] \,\,,
\label{ngeral}
\end{eqnarray}
where   $\gamma$ is the anomalous dimension of the target gluon distribution.    
The main difference among the distinct phenomenological 
models comes from the  predicted behaviour for the anomalous dimension, which 
determines  the  transition from the non-linear to the 
extended geometric scaling regimes, as well as from the extended geometric scaling 
to the DGLAP regime (See e.g. \cite{hdqcd}).  
The current models in the literature consider the general form 
$\gamma = \gamma_s + \Delta \gamma$, where $\gamma_s$ is the anomalous 
dimension at the saturation scale and $\Delta \gamma$ mimics the onset of the 
geometric scaling region and DGLAP regime. 
In what follows we compare the  rcBK \cite{bkrunning}   predictions with those 
of the GBW model \cite{dipolos}, which assumes  
$\gamma (x,\rho^2) = 1$.

In order to estimate  heavy quark production in $pA$ collisions we will use the 
fact  that  color dipoles are  eigenstates 
of the interaction and therefore   the  $q \bar q G$-nucleus interaction can be 
expressed in terms of the eikonalization of the 
cross section on a nucleon target:
\begin{eqnarray}
\sigma^A_{q \bar q G}(x,\rho) = 1 - \exp \left[-\frac{1}{2}  \, 
\sigma^p_{q \bar q G}(x,\rho^2) 
\,T_A(\rb)\right] \,\,,
\label{enenuc}
\end{eqnarray}
where $\sigma^p_{q\bar q G}$ is given by Eq. (\ref{sec2}) and $T_A(\rb)$ is the 
nuclear profile 
function, which is obtained from a 3-parameter Fermi distribution for the nuclear
density normalized to $A$. The above equation, based on the Glauber-Gribov formalism,  
sums up all the  multiple elastic rescattering diagrams of the $q \bar q G$ system
and is justified for large coherence length, where the transverse separation $\rho$ of partons 
in the multiparton Fock state of the projectile becomes a conserved quantity, {\it i.e.} the size 
of the dipole becomes eigenvalue of the scattering matrix. In the next section we will estimate Eq. (\ref{enenuc}) using as 
input the rcBK and GBW models  for the dipole-proton cross section. Moreover, we will compare these predictions with those obtained 
using the CT model generalized to  the nuclear case.

Another issue that we will address is the asymmetry  in heavy quark production in $p \, A$ collisions when we use the color dipole 
formalism. Our treatment of the interaction is not  symmetric under the projectile - target exchange. 
In particular, $p \, A$ and $A \, p$ processes 
allow us to study different physical effects. While $p \, A$ collisions allow to study  non-linear effects in the 
dipole-nucleus interaction, 
the study of $A \, p$ collisions probes the nuclear effects in the  projectile gluon distribution. 
The rapidity distribution in the latter case is 
given by :
\begin{equation}
\frac{d\sigma(A p\rightarrow \{ Q\bar{Q} \}X)}{dy} = x_1G^{A}(x_1,\mu_F^2) \, 
\sigma (Gp \rightarrow \{Q\bar{Q} \}X)\,\,.
\label{dsdyAp}
\end{equation}
As it can be seen, this process can be used to constrain   the magnitude the shadowing and antishadowing effects, which is  still  an open question (For a similar discussion in the Drell-Yan process see Ref. \cite{betemps}).

\section{Results}
\label{resultados}

\begin{figure}[t]
\begin{tabular}{cc}
\includegraphics[scale=0.40]{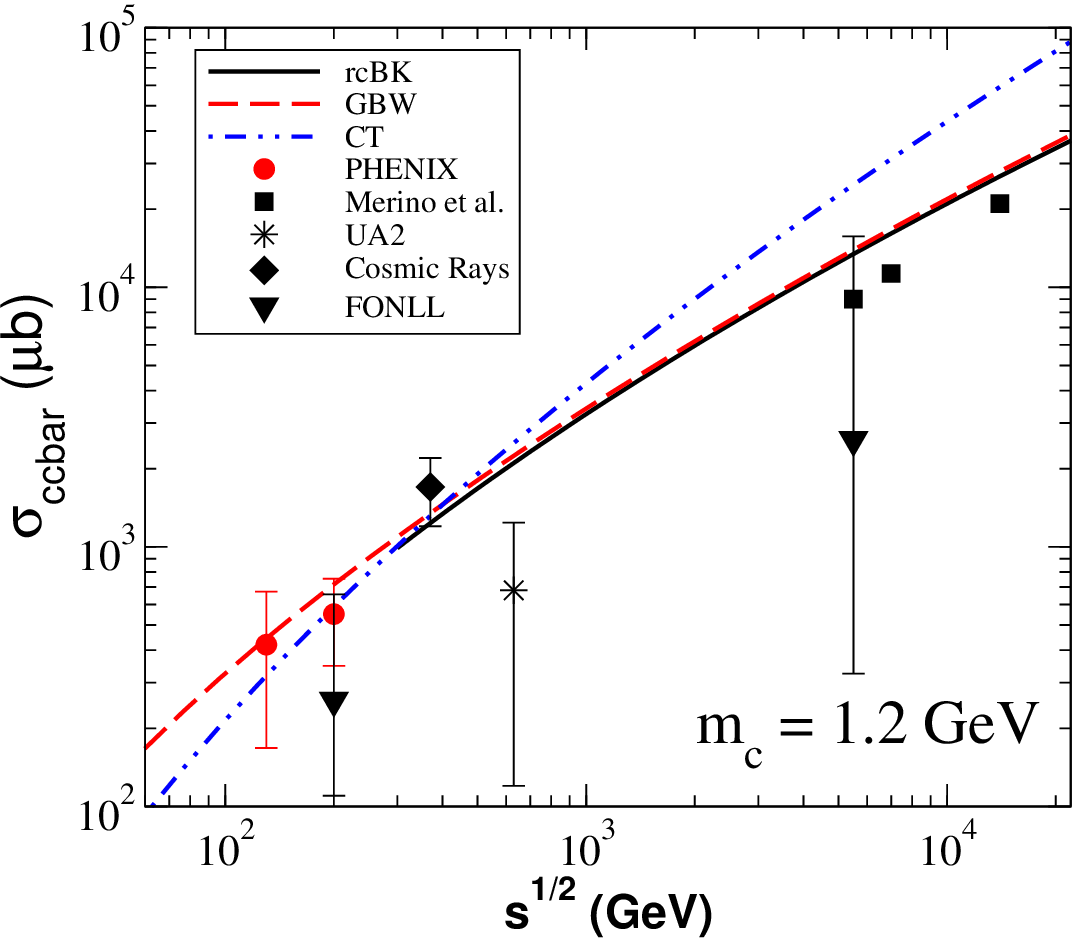}        
\includegraphics[scale=0.40]{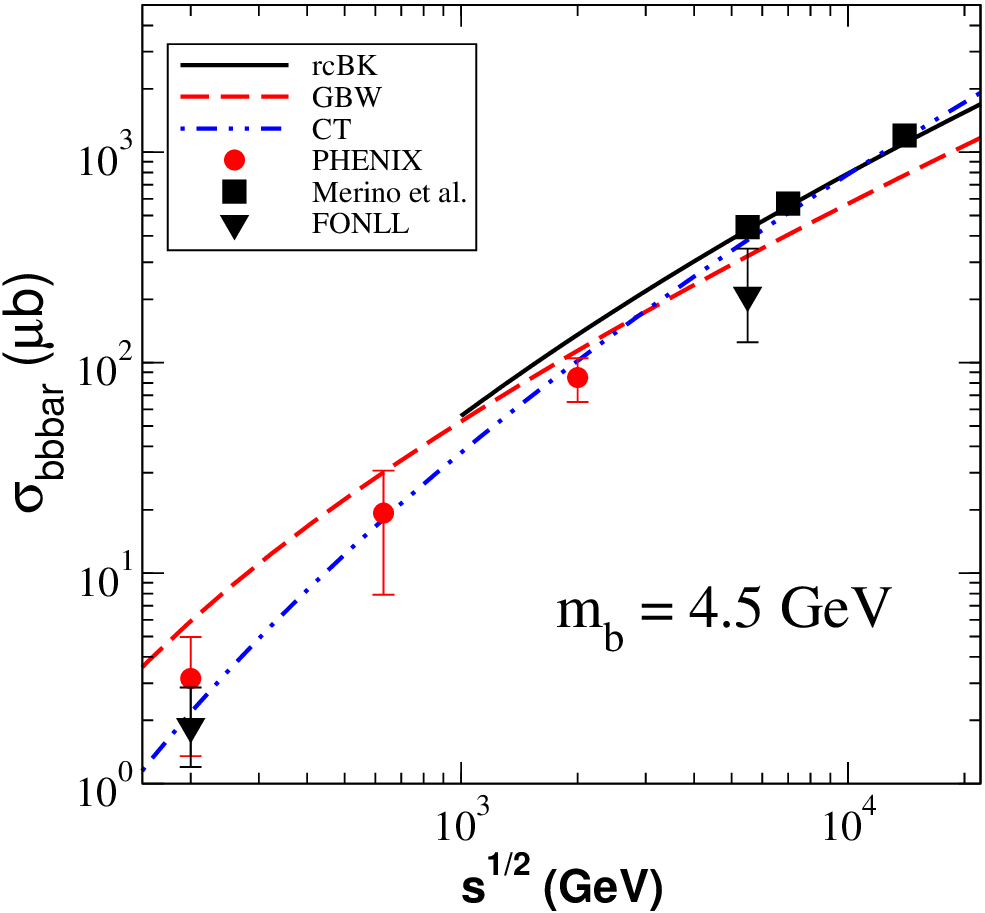}        
\end{tabular}
\caption{(Color online) Total charm (left) and bottom (right) production cross 
sections in proton - proton 
collisions as a function of the c.m.s. energy ($\sqrt{s}$).
Data points from  PHENIX \cite{adcox02,adare06} (circles), from  UA2  \cite{botner} 
(down triangles) and  from  cosmic ray \cite{xu06} (diamonds). 
Theoretical results obtained with  $k_T$  factorization   \cite{merino} 
(squares) and  with  FONLL \cite{ramona} (up triangles).}
\label{fig:1}
\end{figure}

\begin{figure}[t]
\begin{tabular}{cc}
\includegraphics[scale=0.40]{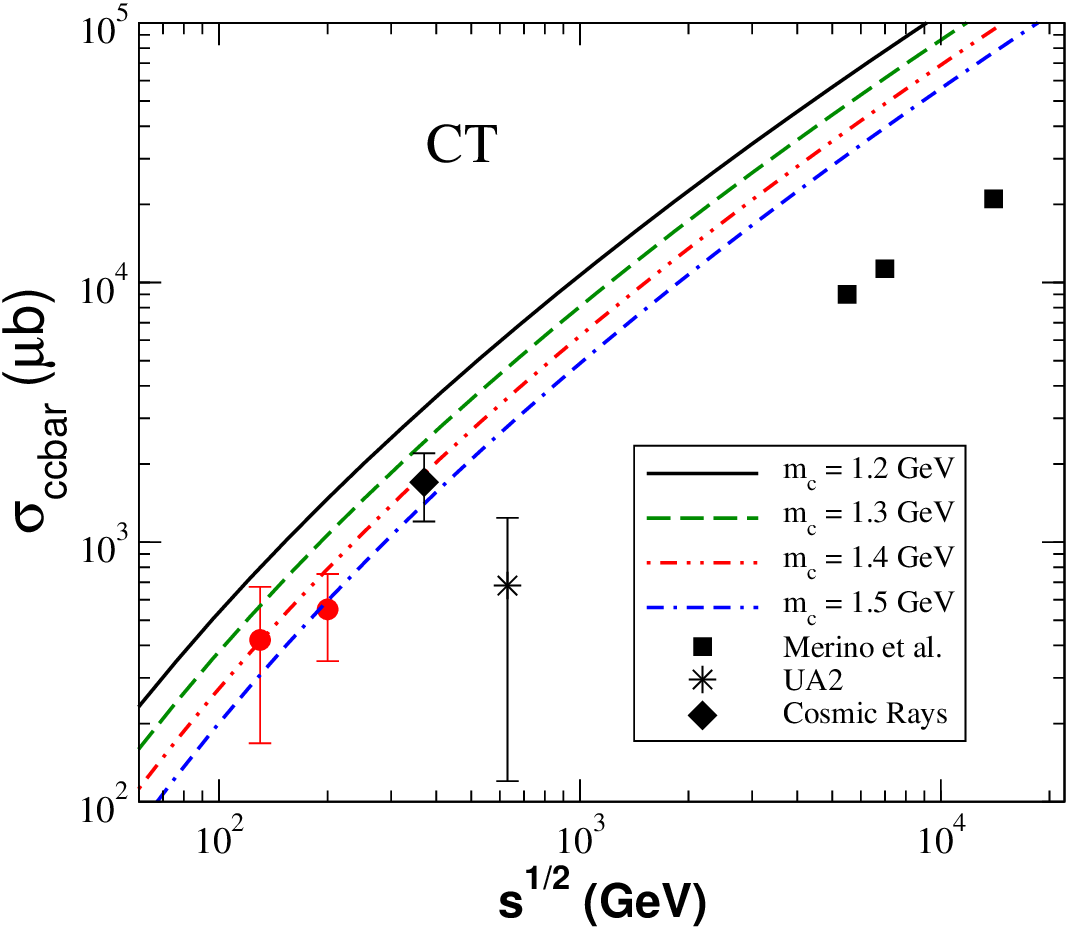}   
\includegraphics[scale=0.40]{heavypA_2a.eps}   
\end{tabular}
\caption{(Color online) Dependence on the mass of the  total charm (left) and bottom (right) production cross 
sections in proton - proton 
collisions as a function of the c.m.s. energy ($\sqrt{s}$)   obtained with
the GBW  dipole cross section.}
\label{fig:2}
\end{figure}

In Fig. \ref{fig:1}   we show the predictions of the color dipole formalism for the 
total heavy quark production cross sections 
in proton - proton collisions as a function of the c.m.s. energy ($\sqrt{s}$)   
using as input  the CT, GBW and rcBK dipole cross 
sections.  The left (right) panels show charm (bottom) cross sections.  As the rcBK 
dipole cross section can be calculated only for small $x$ ($\le 10^{-2}$),   we are not 
able to obtain results for heavy quark
production at the lowest energies.  We compare our results with the experimental 
data obtained by the PHENIX collaboration 
\cite{adcox02,adare06} (circles), by the UA2 Collaboration \cite{botner} (up triangles) 
and with data extracted from  cosmic ray measurements \cite{xu06} (diamonds). We also show the 
theoretical results obtained  with the $k_T$  factorization scheme \cite{merino} 
(squares) and with FONLL \cite{ramona} (down triangles). 
Our results were obtained assuming $m_c = 1.2$  GeV  and $m_b = 4.5$ GeV. 
In the case of the GBW and rcBK predictions  these values of mass allow us to describe 
satisfactorily the experimental data without the inclusion of  a $K$-factor to fit the 
normalization. However, in the CT case it 
is necessary to multiply the  prediction by $K = 0.4$ in order to describe the experimental 
data. This factor becomes equal to one if the mass is increased, since (as in other 
approaches)  there is a  strong dependence of the results on the choice of the heavy 
quark mass. This strong dependence is observed in Fig. \ref{fig:2} for the CT input 
($K = 1$ in this figure) and  similar results are obtained using the GBW and rcBK inputs.  
In what follows we will assume the same values for the heavy quark masses in all 
calculations. Beyond the mass dependence, the heavy quark cross sections are also strongly 
dependent on the renormalization and factorization scales. We postpone the discussion 
about  this dependence  for a future work and  assume  $\mu_R= m_Q$.

From Fig. \ref{fig:1} we can see that in the case of charm production the rcBK and GBW 
predictions are very similar at high energies, while the CT one predicts a stronger growth 
of the cross section with the energy. This difference comes from saturation effects  
which are expected to significantly contribute in the kinematical region in which 
$Q_s^2 > m_Q^2$. On the other hand, in the case of bottom production the rcBK and CT 
predictions are similar and larger than the GBW one. This behavior can be attributed to 
the description of the linear regime which is different  in the rcBK and GBW dipole-proton  
cross sections. In particular, the numerical solutions of the BK equation tend to reach later 
the unitarity limit \cite{vic_joao}. This can be understood looking carefully at the 
integrand of   (\ref{sec1}), which is the product of the  wave functions, containing 
information about the masses, and the dipole cross section. As a function of the dipole 
size, $\rho$, the difference between GBW and rcBK is mostly in the low to intermediate 
$\rho$ region, where the GBW is always below the rcBK dipole cross section. At large 
$\rho$ the two cross sections are close to each other. The squared wave function, i.e., 
$ ({\Psi}^*\Psi)$ given by  (\ref{psi}) has peaks at different locations in $\rho$.  
The $c \bar{c}$ is a larger state  and its wave function peaks at much larger values of 
$\rho$ than the $b \bar{b}$ wave function. In this way it gives a stronger  weight to 
larger $\rho$ where the differences between GBW and rcBK are smaller. A similar effect was
also observed in a previous calculation with  the rcBK dipole cross section. The exclusive 
vector meson production cross sections  change when we change the dipole cross section. One 
of the conclusions of \cite{nos_exclu} was that the difference between results with rcBK and
 bCGC  becomes more pronounced for heavier mesons. 
Our results  suggest that with the total cross section alone we are not able to discriminate 
between the different dynamics contained in the dipole-proton cross sections. On the other 
hand it is reassuring to observe the compatibility of our results both with data and with 
other theoretical estimates.

\begin{figure}[t]
\begin{tabular}{cc}
\includegraphics[scale=0.40]{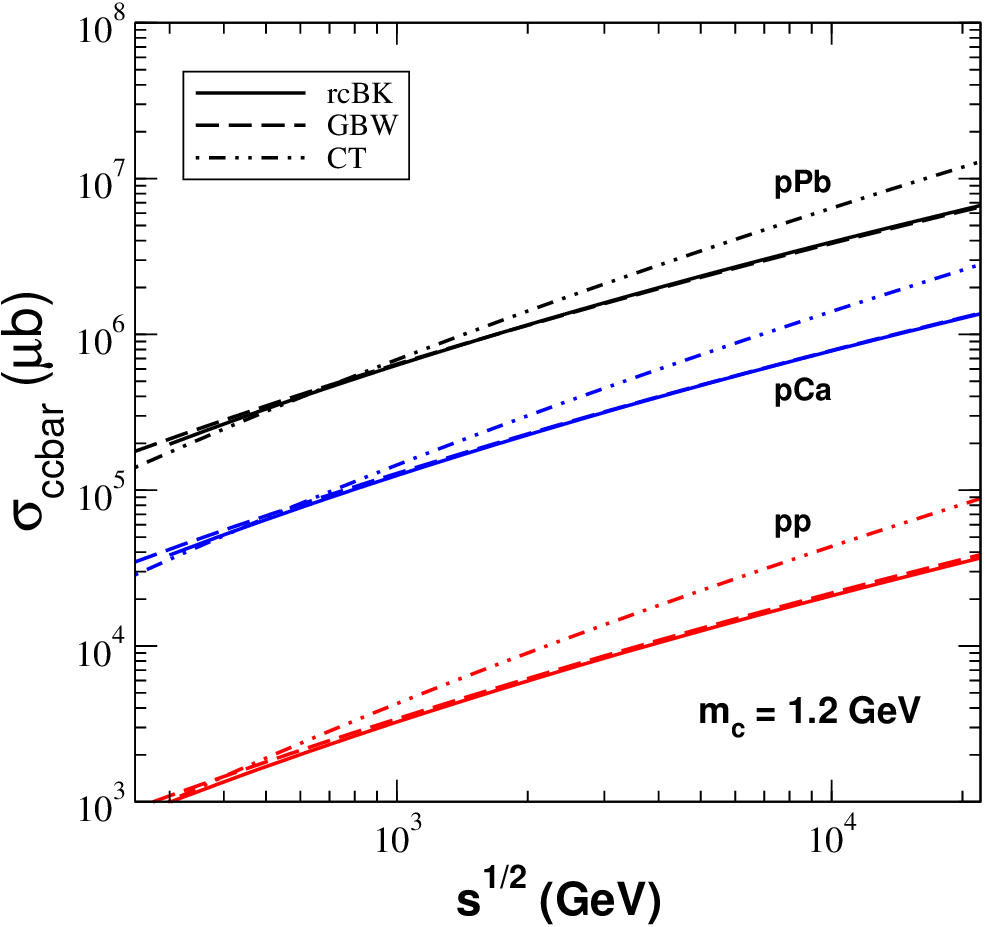}  
\includegraphics[scale=0.40]{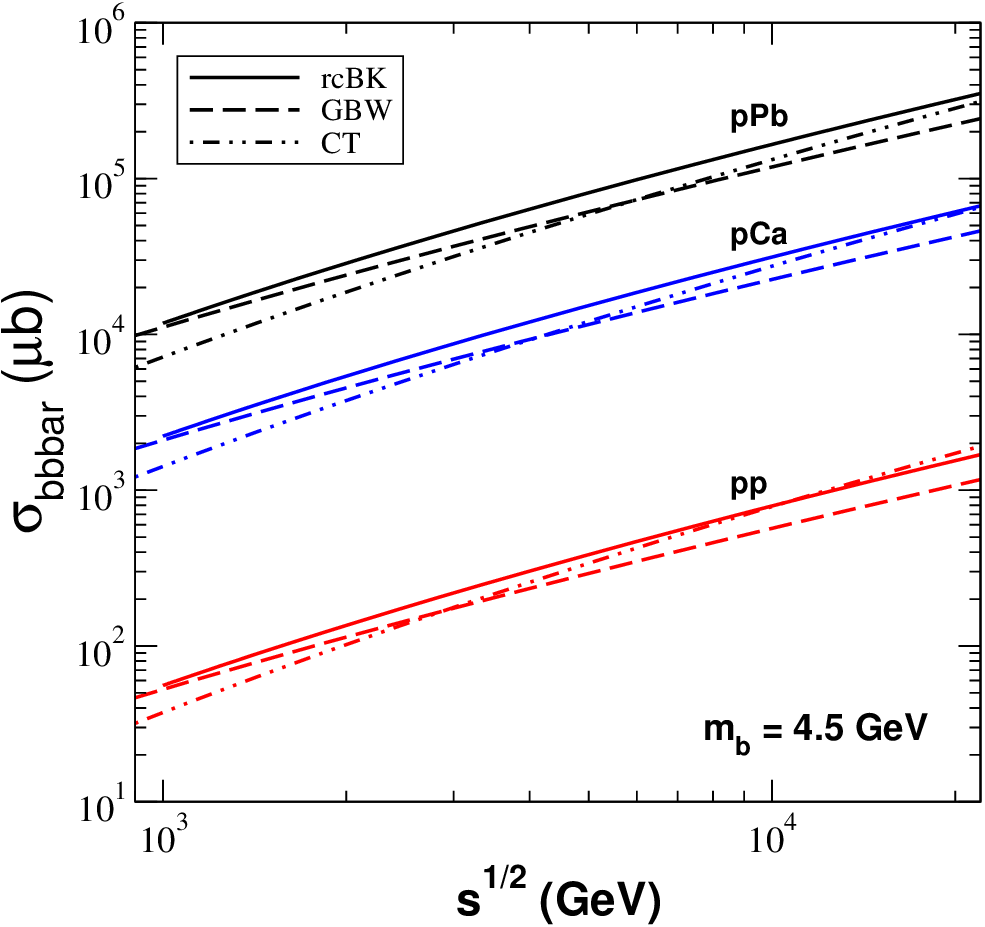}  
\end{tabular}
\caption{(Color online) Target mass dependence  of the  total charm 
(left) and bottom (right) production cross 
sections in $p \, Pb$ and $p \, Ca$ collisions as a function of the c.m.s. 
energy ($\sqrt{s}$). For comparison the $p \, p$ predictions are also show.}
\label{fig:3}
\end{figure}

We address now  heavy quark production in $pA$ collisions. In the absence of nuclear effects, 
the heavy quark cross section would simply scale with the atomic number. Therefore departures
from  the $A$ scaling provide information about nuclear effects. Here we consider the 
influence of saturation and shadowing. Both effects are expected to strongly affect heavy 
quark production at high energies but the  relation between them is not clear. We assume 
two extreme scenarios: the coherent, based on the non-linear dynamics described by CGC 
physics, and the incoherent one, based on the linear DGLAP dynamics. Our goal is to verify 
if it is possible to discriminate these scenarios.   

In Fig. \ref{fig:3} we present our predictions for the heavy quark production in $pA$ 
collisions considering $A = Ca$ and $Pb$. For comparison the $p \, p$ predictions are also 
shown. In the nuclear case and charm production, the GBW and rcBK predictions are very 
similar, as already observed in the proton case. The difference between these predictions 
and those from CT  increases with the the atomic number. This fact can be attributed to 
the growth of the contribution of saturation effects. It is important to remember that the 
nuclear saturation scale is expected to increase with $A^{\frac{1}{3}}$, which implies that 
the kinematical regime where $Q_s^2 > m_Q^2$ is enlarged in  $pA$ collisions at high energies. 
Moreover, the difference between the coherent prediction, obtained using Eq. (\ref{enenuc}), 
and the incoherent one, obtained using the CT model, can be larger if the shadowing effects 
are disregarded in $x \, G_A$. The CT predictions for the nuclear case in  Fig. \ref{fig:3} 
were obtained using the EPS09 \cite{eps09} parametrization of the nuclear effects, which 
predicts a large shadowing effect at small-$x$, implying  a strong reduction of the magnitude 
of the charm cross section. In the bottom case the magnitude of saturation and shadowing 
effects is smaller and therefore the difference between the CT and rcBK predictions is also 
smaller. As in the proton case, the GBW model predicts a distinct behavior in the bottom 
case, that implies a mild growth of the cross section with the energy.

\begin{figure}[t]
\begin{tabular}{cc}
\includegraphics[scale=0.40]{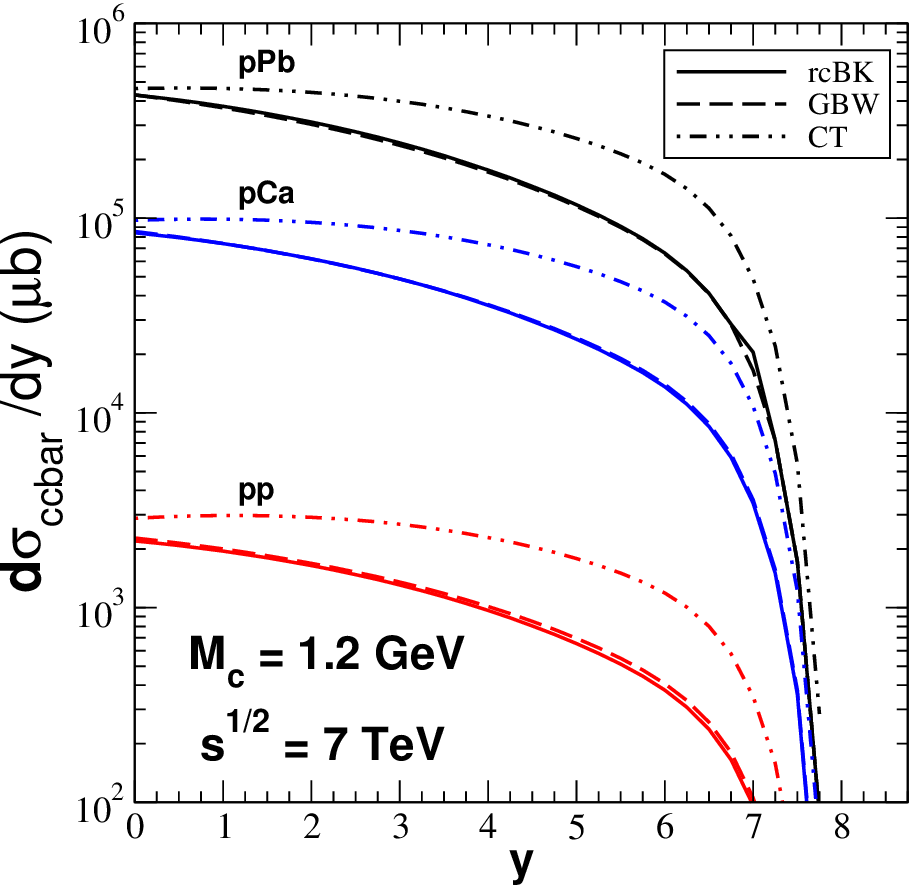} 
\includegraphics[scale=0.40]{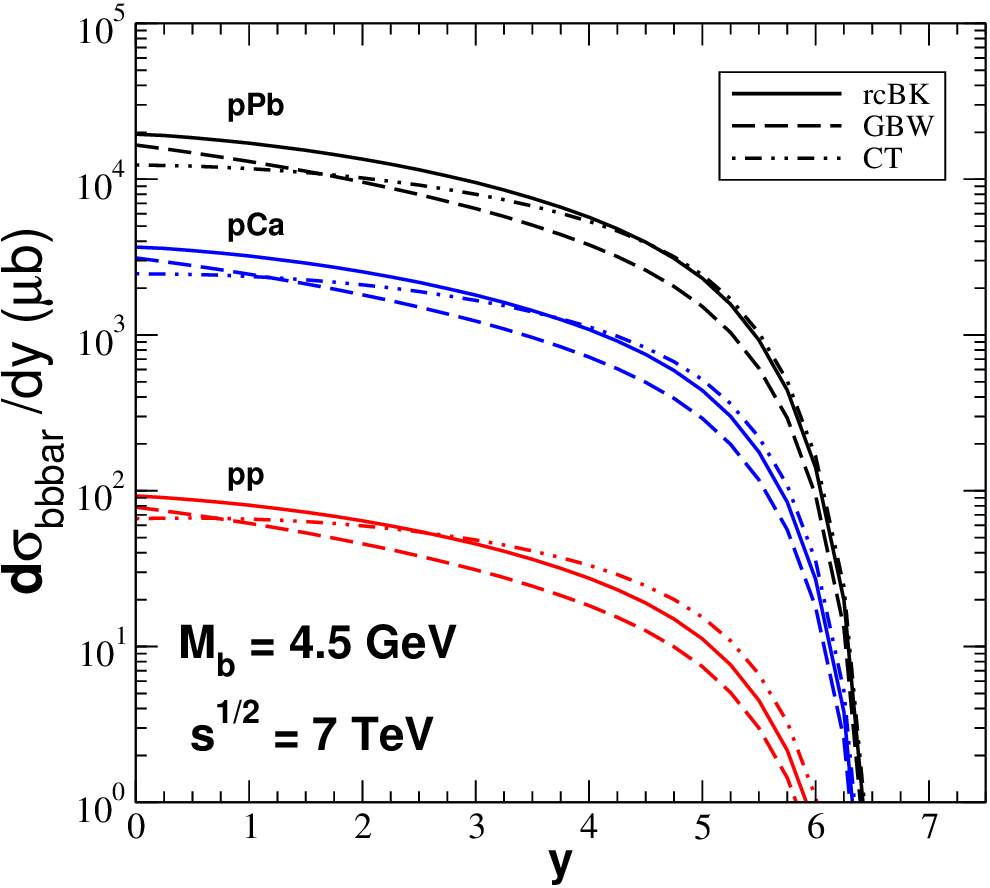} 
\end{tabular}
\caption{(Color online)  Rapidity distributions for charm (left panel) and 
bottom (right panel) production in $p \, p$, $p \, Ca$ and $p \, Pb$ collisions at 
$\sqrt{s} = 7$ TeV.}
\label{fig:4}
\end{figure}

In Fig. \ref{fig:4} we present our predictions for the charm (left panel) and bottom 
(right panel) rapidity distributions in $p \, p$, $p \, Ca$ and $p \, Pb$ collisions. 
For comparison  we use the same value of center-of-mass energy: $\sqrt{s} = 7$ TeV. 
In the charm case, as already observed for the total cross section, the rcBK and GBW 
predictions are very similar and smaller than the CT one, with the difference increasing 
with the rapidity. On the other hand, in the bottom case, the rcBK and CT predictions are 
larger than the GBW one.

\begin{figure}[t]
\begin{tabular}{cc}
\includegraphics[scale=0.40]{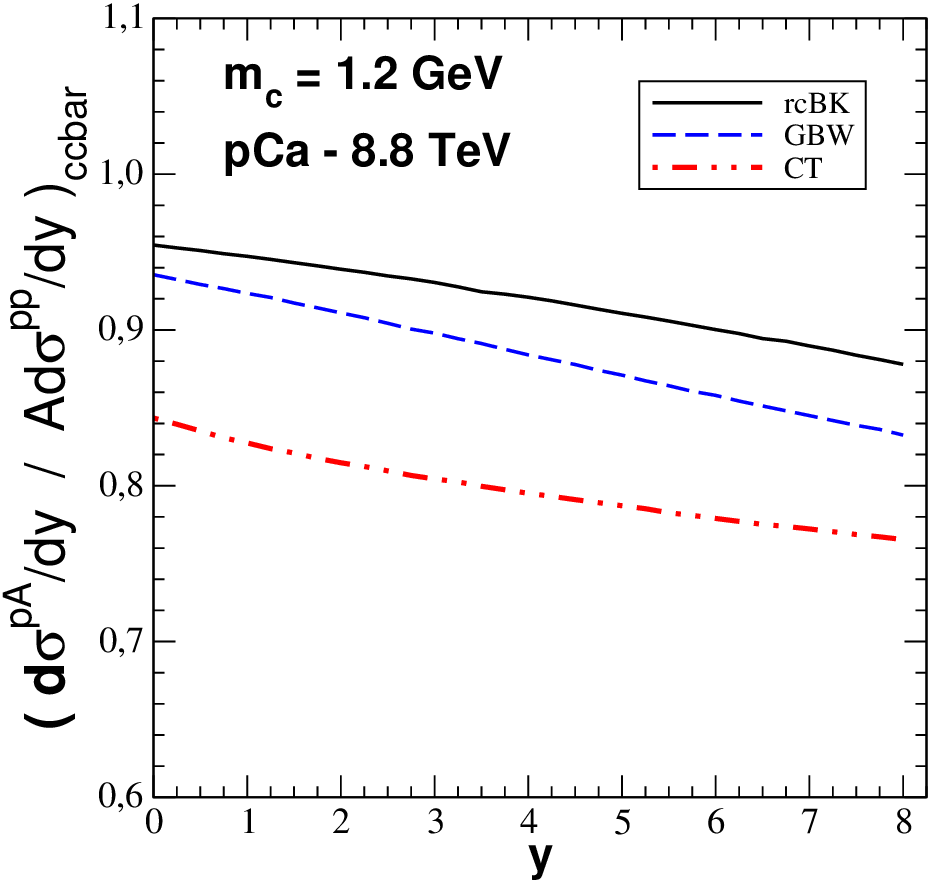} 
\includegraphics[scale=0.40]{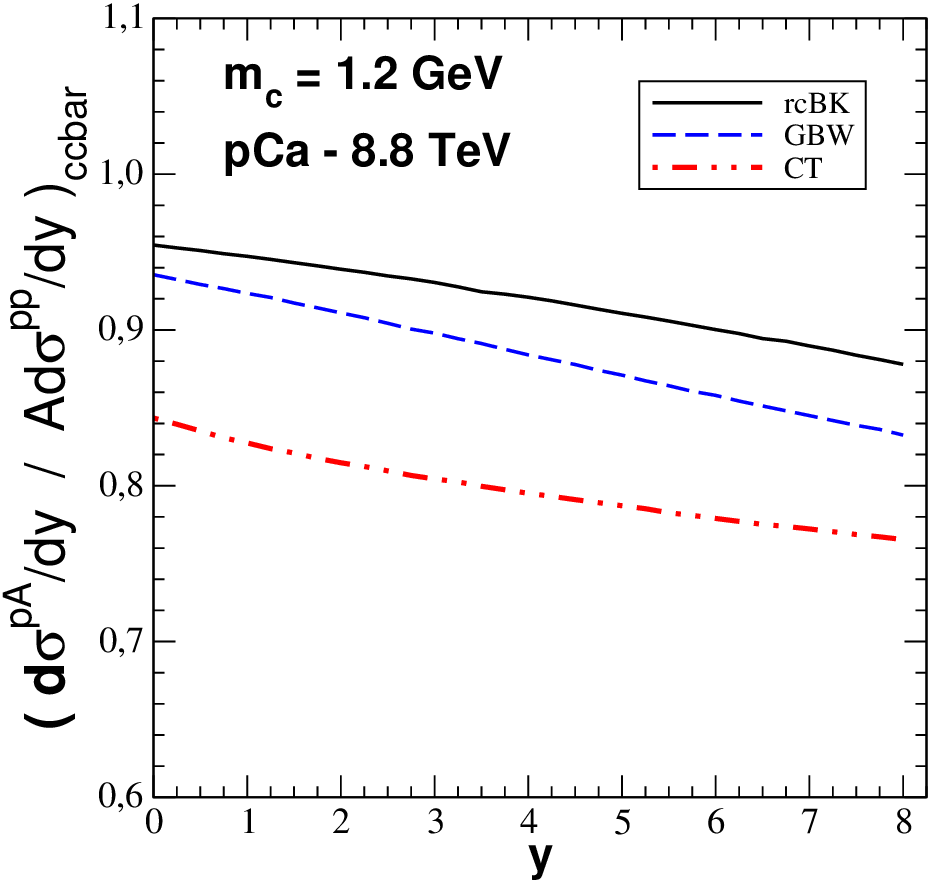} 
\end{tabular}
\caption{(Color online) Rapidity dependence of the normalized ratio between 
charm rapidity distributions  in $p \, Ca$  (left) and $p \, Pb$ (right) collisions at  
$\sqrt{s} = 8.8$ TeV. }
\label{fig:5}
\end{figure}

\begin{figure}[t]
\begin{tabular}{cc}
\includegraphics[scale=0.40]{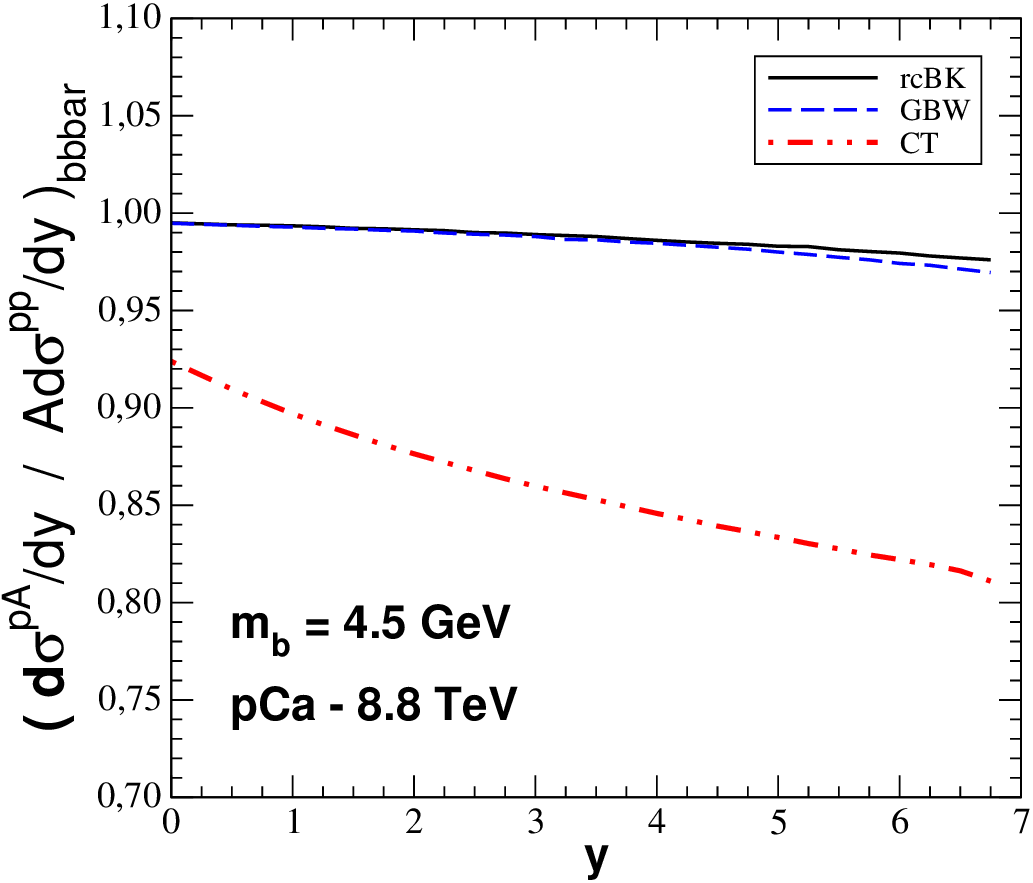} 
\includegraphics[scale=0.40]{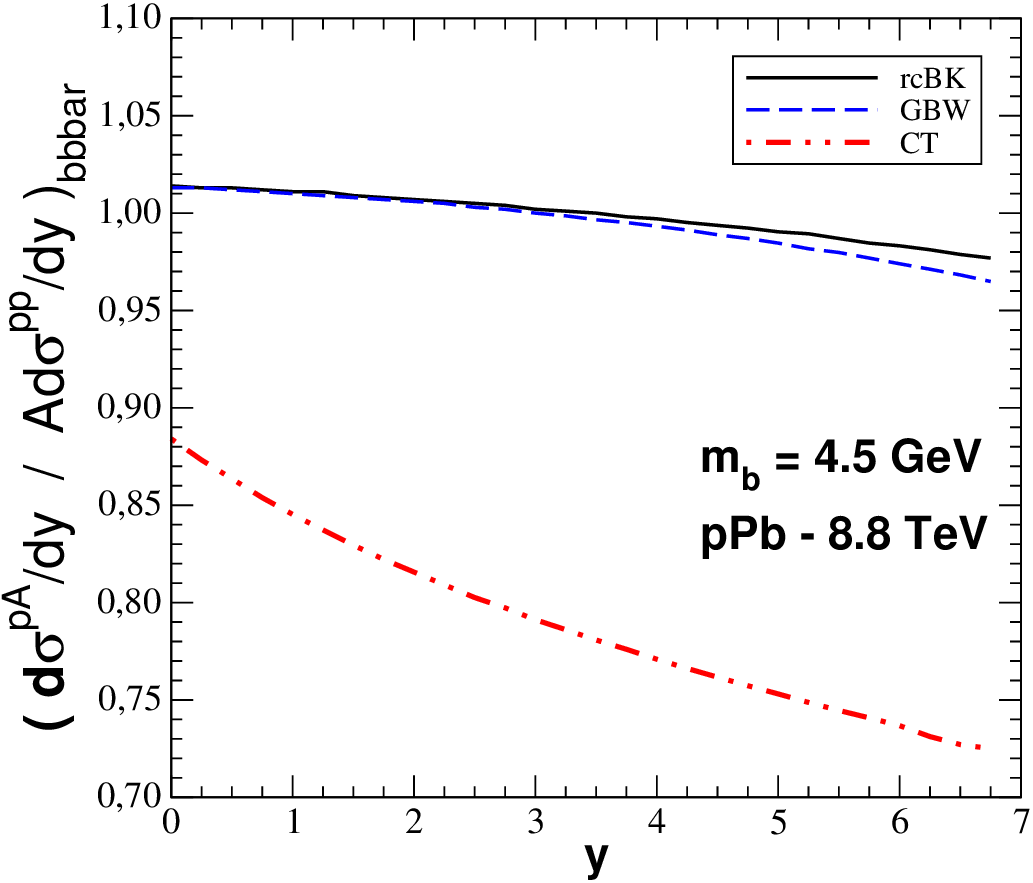} 
\end{tabular}
\caption{(Color online)  Rapidity dependence  of the normalized ratio 
between bottom rapidity distributions  in $p \, Ca$  (left) and $p \, Pb$  (right) 
collisions 
at  $\sqrt{s} = 8.8$ TeV.}
\label{fig:6}
\end{figure}

In order to estimate more precisely the difference between the predictions and the 
magnitude of the effects, in Figs. \ref{fig:5} and \ref{fig:6} we present our predictions 
for the normalized ratio between the rapidity distributions for charm and bottom 
production, respectively. In these figures we choose $\sqrt{s} = 8.8$ TeV, which is the 
expected center-of-mass energy  for $p \, A$ runs at LHC. As expected the ratio diminishes 
at larger $y$ and $A$ and smaller heavy quark mass. In the charm case, the ratio is 
smaller than one for all dipole-target models. While the  rcBK model predicts the  
largest value for the ratio, the CT one predicts the  smallest one. This behavior can be 
related to the magnitude of the shadowing effects present in the EPS09 nuclear gluon 
distribution. In the bottom case, the rcBK and GBW models predict values about one while 
the CT one predicts that the ratio is strongly reduced. Therefore, the behavior of the 
ratio is strongly dependent on the model used to describe the dipole-nucleus interaction.  
If we assume that this interaction is coherent instead of incoherent, smaller values for 
the ratio are predicted.

\begin{figure}[t]
\begin{tabular}{cc}
\includegraphics[scale=0.40]{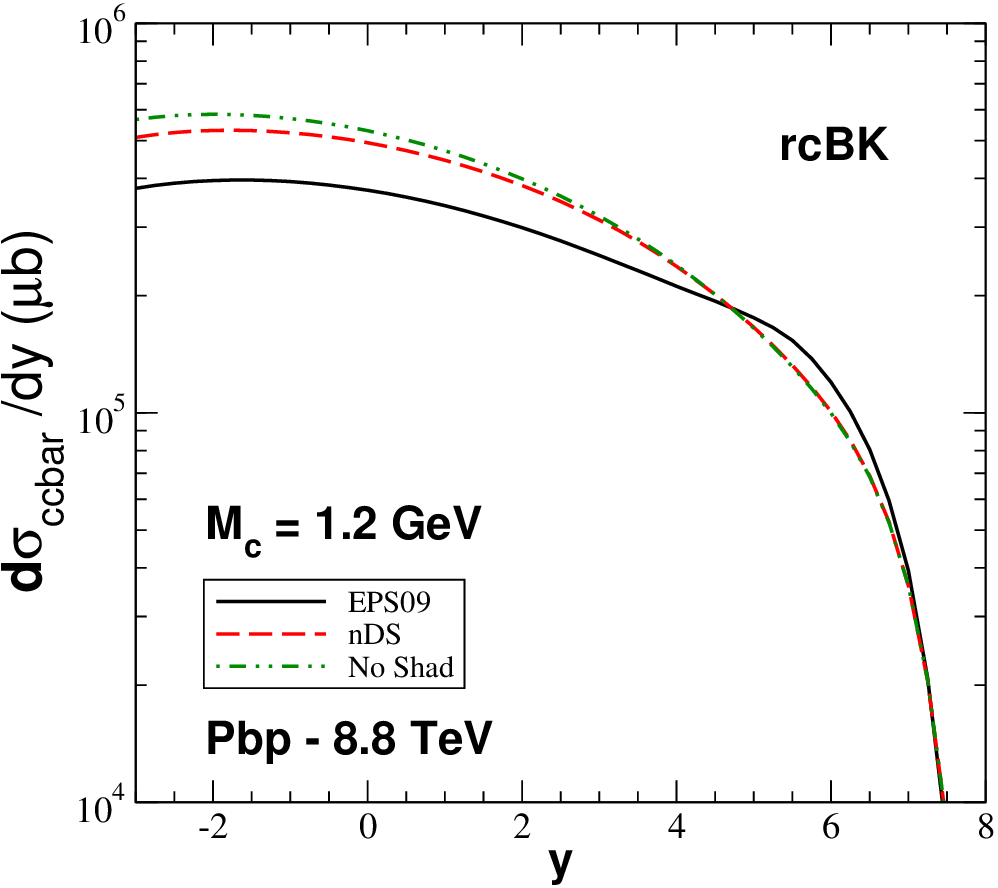}  
\includegraphics[scale=0.40]{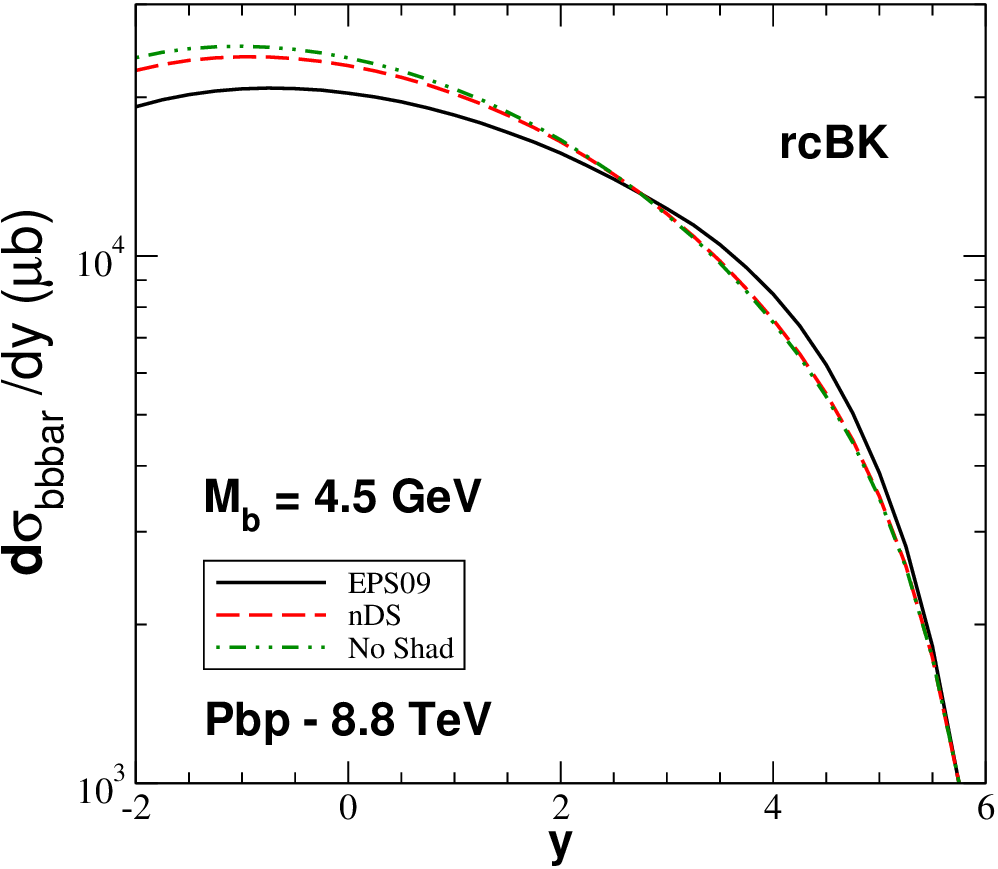}  
\end{tabular}
\caption{(Color online) Rapidity distributions of charm (left) and 
bottom (right ) produced  in $A \, p$ collisions at $\sqrt{s} = 8.8$ TeV.  }
\label{fig:7}
\end{figure}

\begin{figure}[t]
\includegraphics[scale=0.40]{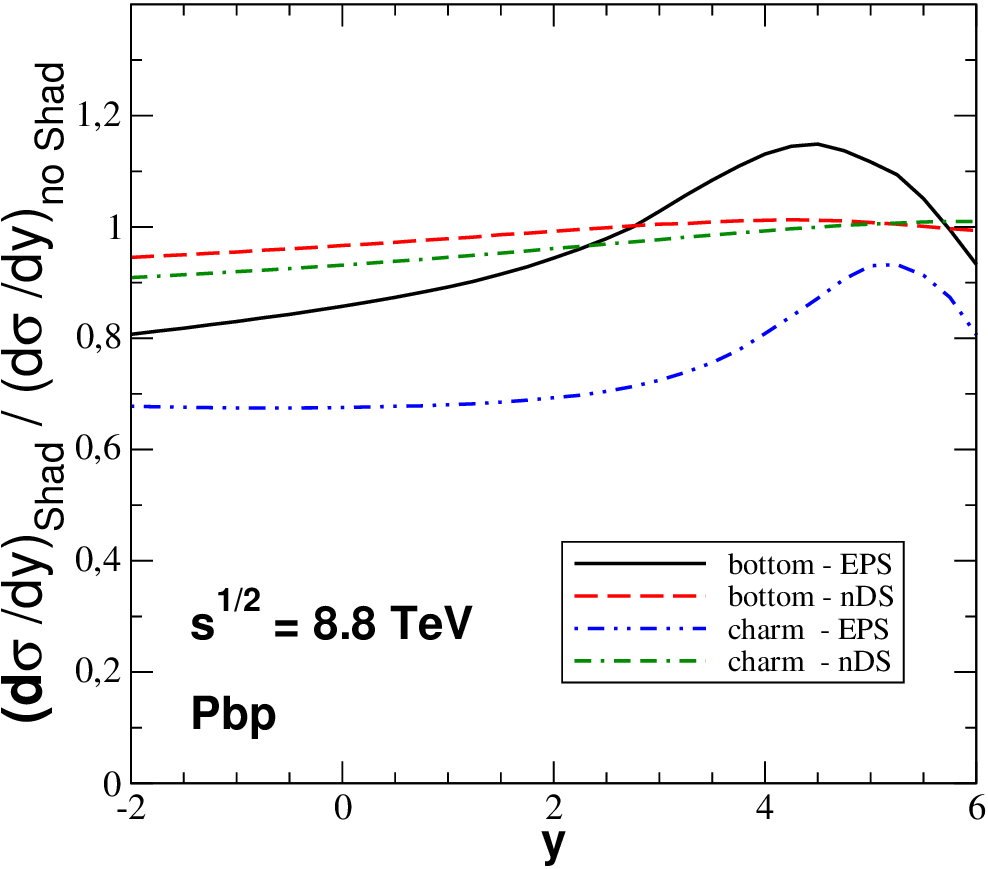}  
\caption{(Color online) Rapidity dependence of the ratio between rapidity 
distributions with and without nuclear effects.}
\label{fig:8}
\end{figure}

As discussed in the previous section, the color dipole formalism for heavy quark production 
in proton-nucleus collisions is asymmetric  under  projectile-target exchange. In what 
follows we estimate the rapidity distribution for charm and bottom production at    
$\sqrt{s} = 8.8$ TeV using Eq. (\ref{dsdyAp}) and assuming that the  gluon distribution 
$x \, G_A$ is given by a parametrization of  the nuclear effects and that the dipole-proton
cross section is given by the rcBK prediction. In Fig. \ref{fig:7} we present our 
predictions considering the EPS09 \cite{eps09} and nDS \cite{nds} parametrizations of the 
nuclear effects. For comparison we also present the prediction obtained disregarding nuclear 
effects. The region of negative rapidity probes small values of $x$ in the nuclear gluon 
distribution [$x_1 = (m_Q/\sqrt{s}) \, e^{+y}$] and consequently the magnitude of shadowing 
effects. At LHC energies, this is also valid at midrapidity. In contrast, at large y we are 
probing  antishadowing. It can be explicitly observed by the coincidence between the 
prediction obtained disregarding the nuclear effects (denoted No Shad in the figure) and the
nDS prediction. One of the basic features of this parametrization  is that 
antishadowing is not present in the gluon distribution. Another feature is the small 
magnitude of shadowing. As observed the nDS and No Shad predictions are similar for the 
bottom production, the nDS one being a little smaller in the case of  charm. In contrast,  
EPS09 predicts a strong reduction at $y < 4.2 \, (2.7)$ and charm (bottom) production. 
This behavior is  more easily observed if we calculate the ratio between the rapidity 
distributions calculated using the EPS09 or nDS and disregarding the nuclear effects. 
The rapidity dependence of this ratio is show in Fig. \ref{fig:8}. This ratio is 
approximately one in the nDS case and present a strong dependence in the EPS09 case.

\section{Summary}
\label{conc}

The description of heavy quark production is one the most  important testing grounds for 
perturbative QCD. Currently, there are several approaches which satisfactorily describe 
the available experimental data. However, it is still an open question whether standard 
pQCD works well at higher energies and in nuclear collisions. In this regime, the 
traditional factorization schemes based on a twist expansion, are expected to breakdown 
due to the presence of saturation effects associated to the high parton density and all 
twists should be resummed. In this paper we have computed  heavy quark production in $pp$ 
and $p \, A$ collisions considering the color dipole formalism, which allows to include  
saturation effects (which are expected to be present at high energies and large nuclei).  
In particular, we analyse the energy dependence of the total cross section and the behaviour 
of the rapidity distribution at fixed energy considering two distinct approaches to treat 
the  dipole - target interaction. One based on the non-linear QCD dynamics and another 
associated to the linear QCD dynamics. Our results shown that the influence of the 
saturation or shadowing effects is different  for  charm and bottom production and therefore   
the simultaneous analyses of the associated observables can be useful to discriminate 
the dynamics. Furthermore, we analyse $A \, p$ collisions as a probe of  shadowing 
effects in  nuclear gluon densities  and demonstrate that the study of  rapidity 
distributions  should allow to constrain their magnitude.



\begin{acknowledgments}
This work was  partially financed by the Brazilian funding
agencies CNPq, CAPES and FAPERGS.
\end{acknowledgments}

\end{document}